\documentclass[10pt]{amsart}

\usepackage{a4wide}

\usepackage{amsfonts}
\usepackage{amssymb}
\usepackage{amsmath}
\usepackage[abbrev]{amsrefs}

\usepackage[latin1]{inputenc}
\usepackage{mathptmx}
\usepackage{dsfont}
\usepackage{mathrsfs}
\usepackage[T1]{fontenc}
\usepackage{calrsfs}
\usepackage{eufrak}
\usepackage{bbm}
\usepackage{xfrac}

\usepackage[colorlinks=true,pdftex]{hyperref}

\usepackage{graphicx}

\newcommand{\del}{\partial}

\newcommand{\ed}{\operatorname{d}}

\newcommand{\expo}{\operatorname{e}}

\newcommand{\idmatrix}{\mathds{1}}

\newcommand{\mcA}{\mathcal{A}}

\newcommand{\mM}{\mathds{M}}

\newcommand{\mR}{\mathds{R}}

\newcommand{\mS}{\mathds{S}}

\newcommand{\mZ}{\mathds{Z}}

\newcommand{\tr}{\operatorname{tr}}

\newcommand{\vphi}{\varphi}

\newcommand{\vth}{\vartheta}

\begin{document}

  
   \title{Blow-up of the non-equivariant 2+1 dimensional wave map}
   
   \address{Department for Mathematics and Statistics\\ University of 
     Otago\\ PO Box 56, Dunedin 9054, New Zealand}
   
   \author{J\"org Frauendiener}
   \author{Ralf Peter}
   
  
   \email{joergf@maths.otago.ac.nz}
   \email{rpeter@maths.otago.ac.nz}
   
   \thanks{RP was supported by a DAAD and a University of Otago Doctoral 
     Scholarship. This research was partially funded by the 
     Royal Society of New Zealand via the Marsden Fund.}


   \begin{abstract}
     It has been known for a
     long time that the equivariant $2+1$ wave map into the 2-sphere blows up if
     the initial data are chosen appropriately. Here, we present
     numerical evidence for the stability of the blow-up phenomenon
     under explicit violations of equivariance.
   \end{abstract}

   
   \maketitle

   
\section{Introduction}
The work presented here, is a continuation of the investigation of the wave map system carried out by the same authors. In~\cite{PF2012} we 
presented our results on the blow-up in the equivariant case which we
obtained by evolution of a $2+1$ code which did not enforce the
equivariance. As we pointed out this is already an indication that the
blow-up phenomenon is stable against perturbations of the size of the
truncation error of the evolution algorithm used. In the present paper
we will give numerical evidence that this is also true for situations
where initial data are used for which equivariance is broken
explicitly already on the level of the initial data.

For this investigation we have used the same setup as
in~\cite{PF2012}. So we will be brief in the description
and for the details refer the reader to that paper.
\subsection{The wave map system}
We use an extrinsic formulation of the $2+1$ wave map, so we
study maps $U$ from $2+1$-dimensional Minkowski space $\mM^{2+1}$ into 
the unit-sphere, embedded into the Euclidean space $\mR^3$
\begin{align*}
  U: \,\, \mM^{2+1} &\longrightarrow \mS^2 \hookrightarrow \mR^3\\
  (x^0,x^1,x^2) &\longmapsto (z^1,z^2,z^3)\,.
\end{align*}
The unit-sphere is described as usual as the zero-set of the
polynomial $\phi(z) = (z^1)^2 + (z^2)^2 + (z^3)^2 - 1 = \delta_{AB}\,
z^Az^B - 1$, where we denoted the Euclidean metric on
$\mR^3$ by $\delta_{AB}$. This implies the restriction 
  \begin{equation}
    \label{eq:constraint}
    \phi(U) = U^A U_A - 1 = 0\,.
  \end{equation}
on the map $U$.

The wave map equations are obtained from an action principle using the action
  \begin{align}\label{eq:action}
    \mcA[U,\del U] 
    = \int_{\mM^{2+1}} \bigl( \del^a U^A \,\del_a U_A + \lambda\phi (U)
    \bigr)\,\ed t\,\ed x\,\ed y.
  \end{align}
Here, $(t,x,y)$ are Cartesian coordinates on $\mM^{2+1}$ and $\lambda$
is a Lagrange multiplier used to implement the constraint~(\ref{eq:constraint}).

Extremising \eqref{eq:action} with respect to $U^A$ and $\lambda$ leads 
  to the Euler-Lagrange equations 
  \begin{align*}
    \square_g U^A - 2\lambda\,U^A &= 0\\
    U^A U_A - 1 &= 0\, ,
  \end{align*}
  where $\square$ is the usual d'Alembert operator $\square = \del_{tt} -
  \del_{xx} -\del_{yy}$. It is possible to 
  eliminate the Lagrange multiplier with the help of the constraint
  equation. However, we choose not do this because in our
  numerical algorithm we solve the constraint and determine the
  Lagrange multiplier at every time-step.
  
  For the sake of clarity, we relabel the component functions of $U$ as
  follows: $u:=U^1$, $v:=U^2$ and $w:=U^3$. Then, we can 
  write the wave map system in the form:
  \begin{align}
  \left\{
  \begin{aligned}\label{eq:WM_extr_2+1_Cartesian}
    \ddot{u} - \del_{xx}u - \del_{yy}u - 2\lambda u &= 0\,\,\\
    \ddot{v} - \del_{xx}v - \del_{yy}v - 2\lambda v &= 0\\
    \ddot{w} - \del_{xx}w - \del_{yy}w - 2\lambda w &= 0
  \end{aligned}\right.\\
    \label{eq:constraint_2+1}
    \phi(u,v,w) = u^2 + v^2 + w^2 - 1 &= 0\,.
  \end{align}
  
  This system has two non-trivial static solutions $U_{\text{S}}$
  \begin{align}\label{eq:static_solution}
    u_{\text{S}}(x,y) = \frac{2x}{1 + x^2 + y^2}\,,\quad
    v_{\text{S}}(x,y) = \frac{2y}{1 + x^2 + y^2}\,,\quad
    w_{\text{S}}(x,y) = \pm\,\frac{1 - x^2 - y^2}{1 + x^2 + y^2}\,.
  \end{align}
  They describe the inverse of the stereographic projection to the sphere from 
  the north- resp.\ south pole.
\subsection{Blow-up dynamics}\label{sec:bl-dynamics}
  The key feature to investigate the blow-up of the $2+1$ dimensional wave map 
  system is the scaling invariance of the equations (holds in all dimensions) 
  and 
  the energy (only in $2+1$ dimensions). This means that the equations as well 
  as the energy are invariant under the transformation
  \begin{align*}
    (t,x,y) \longrightarrow (st,sx,sy) \quad \text{with} \quad s\in\mR\,.
  \end{align*}
  Due to the fact that the energy is also scaling invariant, the $2+1$ 
  dimensional case is called the energy critical case.

%
  The first numerical results on the blow-up of the equivariant system
  were obtained by Bizo\'n et. al. in \cite{BCT2001}. The following
  three observations were made: First, when the energy of the initial
  data is too large then a singularity will form. Later, Sterbenz and
  Tataru~\cite{ST2010_2} specified in more detail under what
  conditions the $2+1$ wave map with the 2-sphere as target, has
  non-singular solutions.
  
  Second, close to the blow-up it is possible to rescale the dynamic
  solution $U$ of \eqref{eq:WM_extr_2+1_Cartesian} so that it
  approximates (in an appropriate Sobolev space) the static solution
  \eqref{eq:static_solution}:
  \begin{align}\label{eq:blow-up_behaviour}
    \lim_{t \nearrow T} U(t,s(t)x,s(t)y) = U_{\text{S}}(x,y)
  \end{align}
  where $T$ is the blow-up time and $s(t)$ is the so called scaling function. 
  The blow-up respectively the singularity formation appears as a shrinking of 
  the rescaled static solution \eqref{eq:static_solution}. This result was 
  proven by Struwe~\cite{Struwe2003_2}. In the same article was also shown 
  that the existence of a non-trivial static solution is necessary for 
  singularity formation.
  
  The scaling function $s(t)$ can be used to detect how the
  singularity formation proceeds. This was used in \cite{BCT2001} and
  \cite{PF2012} for the numerical investigation of the
  blow-up. Bizo\'n et. al. stated two properties for the scaling
  function: $s(t)>0$ for $t<T$ (there is no solution for $t>T$
  anymore) and $s (t)\searrow 0$ for $t \nearrow T$. Rapha\"el and
  Rodnianski~\cite{RR2009} as well as Ovchinnikov and
  Sigal~\cite{OS2011} presented a detailed work on the blow-up
  dynamics. In both articles, an analytic form for the scaling
  function $s(t)$ was obtained. Ovchinnikov and Sigal were able to
  reduce the number of free parameters and therefore could give a more
  precise description of the scaling function $s(t)$.
  
  The third observation was that towards the blow-up the local kinetic
  energy at the point of the singularity formation goes to zero and
  the local potential energy approaches the value $4\pi$, which is the
  energy of the static solution \eqref{eq:static_solution}. 
  
  Those results were confirmed by Isenberg and Liebling~\cite{IL2002}. In our 
  previous work \cite{PF2012} we also observed the expected blow-up behaviour. 
  In addition we showed that the blow-up is stable under perturbations with a 
  magnitude of the truncation error of the numerical scheme.
\subsection{Numerical setup}
The numerical method which will be used to solve the equations
\eqref{eq:WM_extr_2+1_Cartesian} and \eqref{eq:constraint_2+1} is the
same as given in~\cite{PF2012}. Therefore, we will only briefly
outline the most important points here. We discretise the spatial
derivatives in the action functional using fourth order centred finite
differences. This yields a semi-discrete action for finitely many
degrees of freedom. The Euler-Lagrange equations for this action gives
Hamiltonian equations of motion which are symplectic by construction,
i.e., they preserve the canonical symplectic form of classical
mechanics. The constraint \eqref{eq:constraint_2+1} results in as many
holonomic constraints as there are points on the numerical grid. In
this way we obtain a Hamiltonian system with holonomic constraints.
Our time integration method takes advantage of these properties: we
use the Rattle method \cite{Andersen1983}, a symplectic integrator for
Hamiltonian systems with holonomic constraints.

As described in \cite{PF2012} we use the unit square $\Omega =
[0,1]\times [0,1]$ as our domain of integration with homogeneous
Neumann boundary conditions at the outer boundary, i.e., on the sides
with $x=1$ and $y=1$, respectively. The other sides we regard as lines
of symmetry, thus effectively enlarging the domain of computation to
the square $[-1,1]\times [-1,1]$. We use equal resolutions in both
directions, i.e., $\Delta x = \Delta y = 1/(N-1)$ on a grid with $N$
grid points in each direction.
  
\subsection{Initial data}
  To guarantee that the constraint equation \eqref{eq:constraint_2+1} is 
  satisfied, the initial data are chosen as
  \begin{align*}
    u_0(0,x,y) &= \sin(\vth_0(r,\sigma))\,\cos(\vphi_0(\sigma))\\
    v_0(0,x,y) &= \sin(\vth_0(r,\sigma))\,\sin(\vphi_0(\sigma))\\
    w_0(0,x,y) &= \cos(\vth_0(r,\sigma)),
  \end{align*}
  where $r$ and $\sigma$ are polar coordinates in $\mR^2$
  related to the Cartesian coordinates $x$ and $y$ in the usual way
  \[
      x = r\,\cos(\sigma),\quad      y = r\,\sin(\sigma).
  \]

  `Equivariance' means that rotations in the $(x,y)$-plane in
  $\mM^{2+1}$ are mapped to rotations around the $z$-axis in $\mR^3$
  and is reflected in this parametrisation of the initial data as the
  requirements that
  \[
  \vth_0(r,\sigma) = \vth_0(r), \quad \vphi_0(\sigma) = k \sigma \quad
  \text{ with } k \in \mZ. 
  \]

\begin{figure}[htb]
     \centering
     \includegraphics{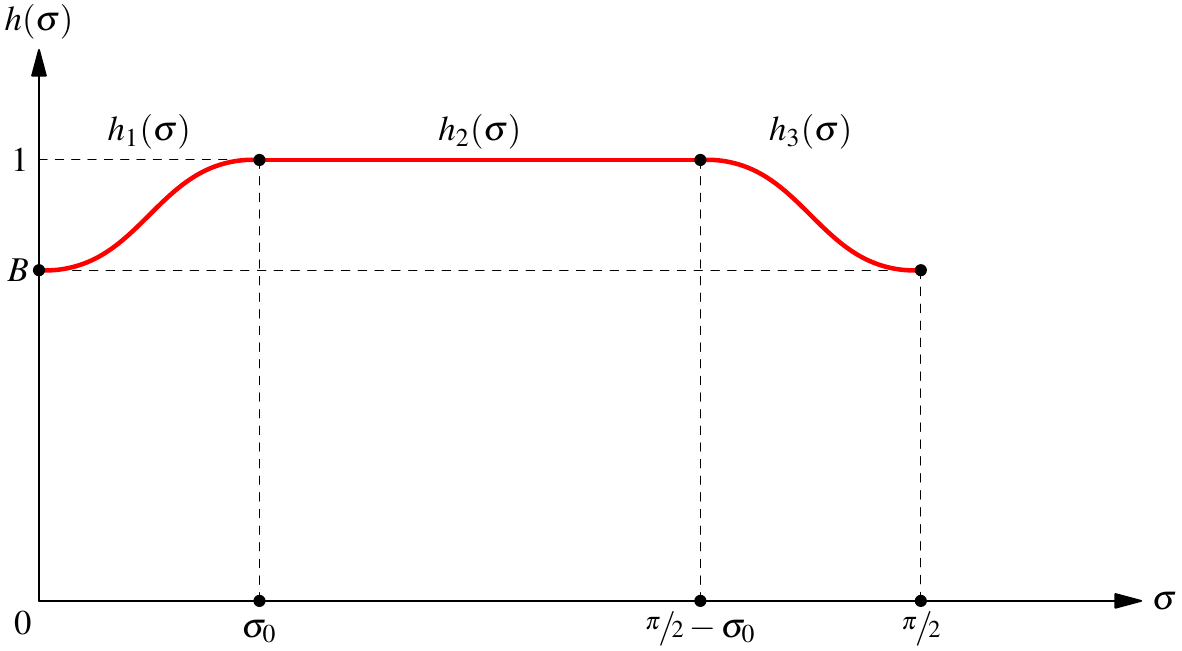}
     \caption{The angular perturbation function $h(\sigma)$.}
     \label{fig:AngularPerturbation}
  \end{figure}

  In our choice of initial data we explicitly break the equivariance
  by making $\vth_0$ depend on $\sigma$ but we keep the reflection
  symmetry across the lines $x=0$ resp. $y=0$ discussed above. The
  function $\vth_0(r,\sigma)$ is defined as follows
  \begin{align*}
    \vth_0(r,\sigma) &= 
    \begin{cases}
      A\,g(r)\,h(\sigma)\quad \text{for} \quad r\in [r_1,r_2]\\
      0\quad\text{otherwise}
    \end{cases}\\
    g(r) &= \left[4\frac{(r - r_1)}{(r_2 - r_1)}\frac{(r_2 - r)}{(r_2 - r_1)}\right]^4\\
    h(\sigma) &= \begin{cases}
                  h_0(\sigma) \quad &\text{for} \quad \sigma \in [0,\sigma_0]\\
                  1 \quad &\text{for} \quad \sigma \in \left
                  [\sigma_0,\frac{\pi}{2} - \sigma_0\right]\\
                  h_0(\frac\pi2 - \sigma) \quad &\text{for} \quad \sigma \in \left[\frac
                  {\pi}{2} - \sigma_0,\frac{\pi}{2}\right]
                \end{cases}
  \end{align*}
  where $h_0(\sigma)$ is a monotonically increasing function with
  $h_0(0)=B$ and $h_0(\sigma_0)=1$. It is chosen in such a way that $h$
  is at least $\mathcal{C}^4$, see Fig.~\ref{fig:AngularPerturbation}.
  The function $h(\sigma)$ describes the deviation from equivariance,
  which would correspond to $h(\sigma)=1$. The parameter $B$ measures
  the strength of the deviation ($B=1$ corresponds to the equivariant
  case).
  We choose $\vphi_0(\sigma) = \sigma$ as in the equivariant case with
  homotopy index $k=1$.
  
  The initial data for the velocities are chosen as
  \begin{align*}
     \dot{u}(0,x,y) &= \del_r u = \cos \vth_0(r,\sigma)\,\vth_0'(r,\sigma)\,\frac{x}{r}
     \\
     \dot{v}(0,x,y) &= \del_r v = \cos \vth_0(r,\sigma)\,\vth_0'(r,\sigma)\,\frac{y}{r}
     \\
     \dot{w}(0,x,y) &= \del_r w = - \sin \vth_0(r,\sigma)\,\vth_0'(r,\sigma)
     \vphantom{\frac{y}{r}}\\
  \end{align*}
  where $\vth_0'(r)$ denotes the derivative of $\vth_0(r)$ with respect to its 
  argument. These initial data describe a ring-shaped bump in the 
  $xy$-plane around the origin. The choice of the velocity initial data results 
  in a shrinking of this ring towards the origin. After the function $w(t,x,y)$ 
  reached its minimum close to the origin, the wave packet expands again.
\section{Scaling function}
  As described above the blow-up dynamics is captured by the scaling
  function $s(t)$ between the solution and the approximated static
  solution. In~\cite{PF2012} we described how we determine this
  function in the equivariant case: since $w$ is an axisymmetric
  function $w(t,r)$ in that case we determine its second derivative with
  respect to $r$ at $r=0$ at every time $t$ and find $s(t)$ as the
  appropriate factor between this and the second radial derivative of
  the static solution. 

  When equivariance is broken then $w$ is no longer axisymmetric and
  we need to extract the scaling function from the full matrix of
  second derivatives, the Hessian, of $w$ at the origin. The Hessian
  $H_{\text{S}}(t,x,y)$ of the rescaled static solution
  \begin{align*}
     w^s_{\text{S}}(t,x,y) := w_{\text{S}}\left(\sfrac{r(x,y)}{s(t)}\right) 
     = \frac{1 - \left(\sfrac{r(x,y)}{s(t)}\right)^2}{1 + \left(\sfrac{r(x,y)}{s
     (t)}\right)^2}\,.
  \end{align*}
  at the origin is proportional to the identity matrix:
  \begin{align*}
    H_{\text{S}}(t,0,0) = - \frac{4}{s^2(t)}\,\idmatrix_2\,.
  \end{align*}
  If the blow-up dynamics in the non-equivariant case is similar to
  the equivariant case then close to blow-up the off-diagonal terms of
  the Hessian of the solution will be small compared to the diagonal
  terms and we can extract the scaling function $s(t)$ simply from the
  trace of the Hessian using
\[
    \tr H_{\text{S}}(t,0,0) = - \frac{8}{s^2(t)}.
\]
 Alternatively, we could find $s(t)$ also by
  taking the determinant of the Hessian using
\[
    \det H_{\text{S}}(t,0,0) = \frac{16}{s^4(t)}\,.
\]  
  Geometrically this just means that we take either the mean curvature or the Gauss
  curvature at the origin of the surface defined by the graph of $w$
  as the indicator for the scaling function. It turned out that there
  are no essential differences so we used the Gauss curvature throughout.

\section{Blow-up results}
In Subsection \ref{sec:bl-dynamics} the dynamics of the wave packet
was described. Now we want to give some more details about the
behaviour of the solution for initial data with large energy. If the
energy of the initial data is large enough one expects a singularity
formation as it is presented in \cite{BCT2001} and \cite{PF2012}. In
our setup, we interpret a change in the behaviour of $w(t,x,y)$ at the
origin as the appearance of the blow-up. We have specified the initial
data so that there remains a residual symmetry, namely the reflection
symmetry with respect to both coordinate axes. This symmetry has the
consequence that the value of the function $w$ should remain constant,
i.e., with our initial data $w(t,0,0)=1$ throughout the
evolution. However, if the energy is too large the numerical solution
suddenly changes $w(t,0,0) = -1$. From a geometrical point of view the solution switches from the stereographic projection from the
south pole to the projection from the north pole, i.e., it suddenly
approximates \emph{the other static solution}. The reason for this
seems to be due to the numerical
method we are using. The Rattle method for the time integration always
forces the solution onto the constraint manifold. For large energies
it becomes increasingly difficult for the iterative projection
algorithm to find a solution. It seems it is somewhat easier for the
system to flip to the other solution.

  Using the previously introduced methods, we are now able to analyse the blow-up 
  dynamics and singularity formation in the non-equivariant case.
  This will be analogous to the equivariant procedure 
  presented in \cite{PF2012}. For these simulations 
  the value of the deviation from the equivariance $B$ is fixed and the 
  amplitude $A$ is increased towards the critical value. As the
  indicator for the presence of the blow-up singularity we take the
  above mentioned flip from one static solution to the other.
  
\begin{figure}
    \centering
    \includegraphics{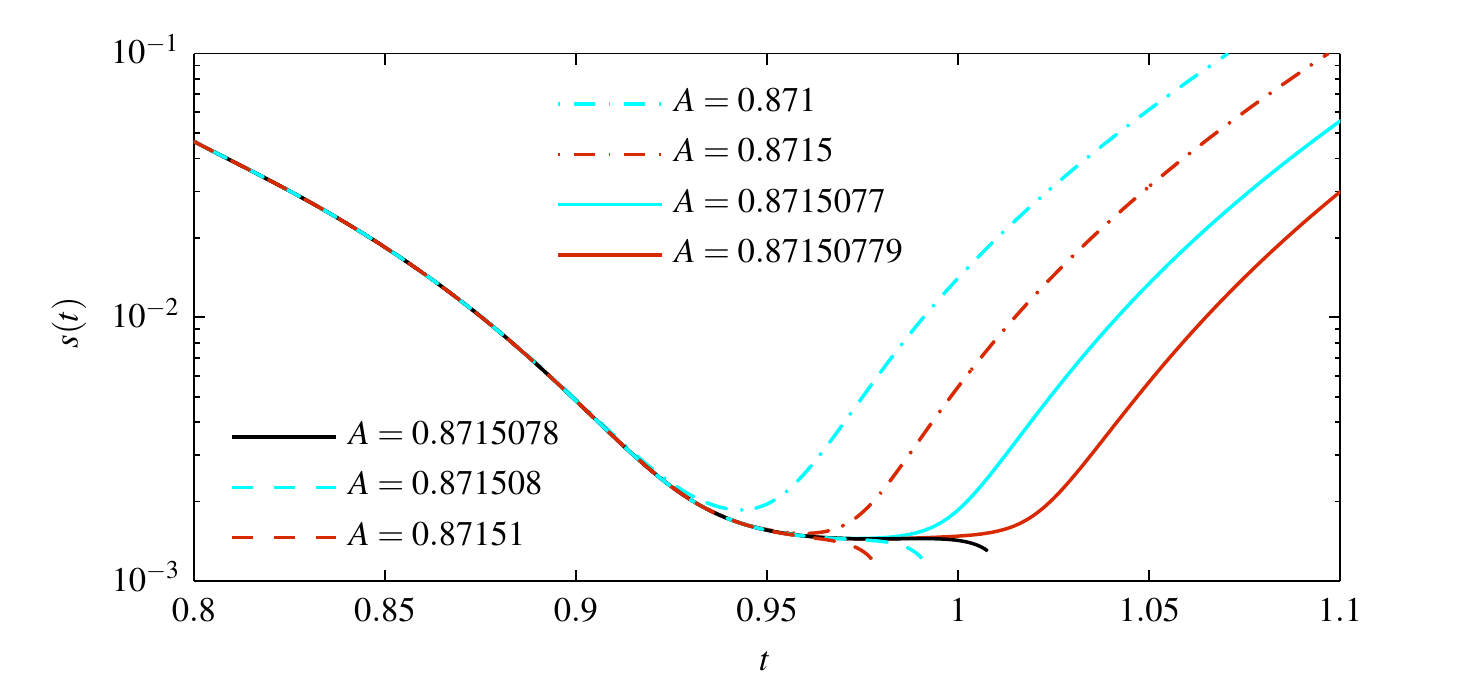}
    \caption{The scaling function $s(t)$ for different parameter values $A$ and 
       $B = 0.8$. The value $A^* = 0.87150780$ is the critical amplitude. 
       Computed with $N = 1281$.}
    \label{fig:ScalingFunction_N1281_TC1_NonEquiv}
  \end{figure}

  The first step in the analysis of the blow-up is the determination of the 
  critical amplitude $A^*$.
  Figure \ref{fig:ScalingFunction_N1281_TC1_NonEquiv} shows the
  scaling function for different values of the amplitude $A$, where
  the parameter $B$ is fixed to $B=0.8$ and the numerical resolution
  is $N = 1281$.  The qualitative behaviour is the same as in the
  equivariant case. An increase of the amplitude towards the critical
  value $A^* \approx 0.87150780$ leads to an increase in the time the
  system remains in the quasi-static state. The appearance of this
  quasi-static, hovering state is due to the limited spatial
  resolution. If the number of grid points for the simulations is
  increased, the critical behaviour moves to higher values of the
  amplitude and reduces the duration of the hovering state. Therefore,
  our calculations are limited by the spatial resolution.
  
\begin{figure}
    \centering
    \includegraphics{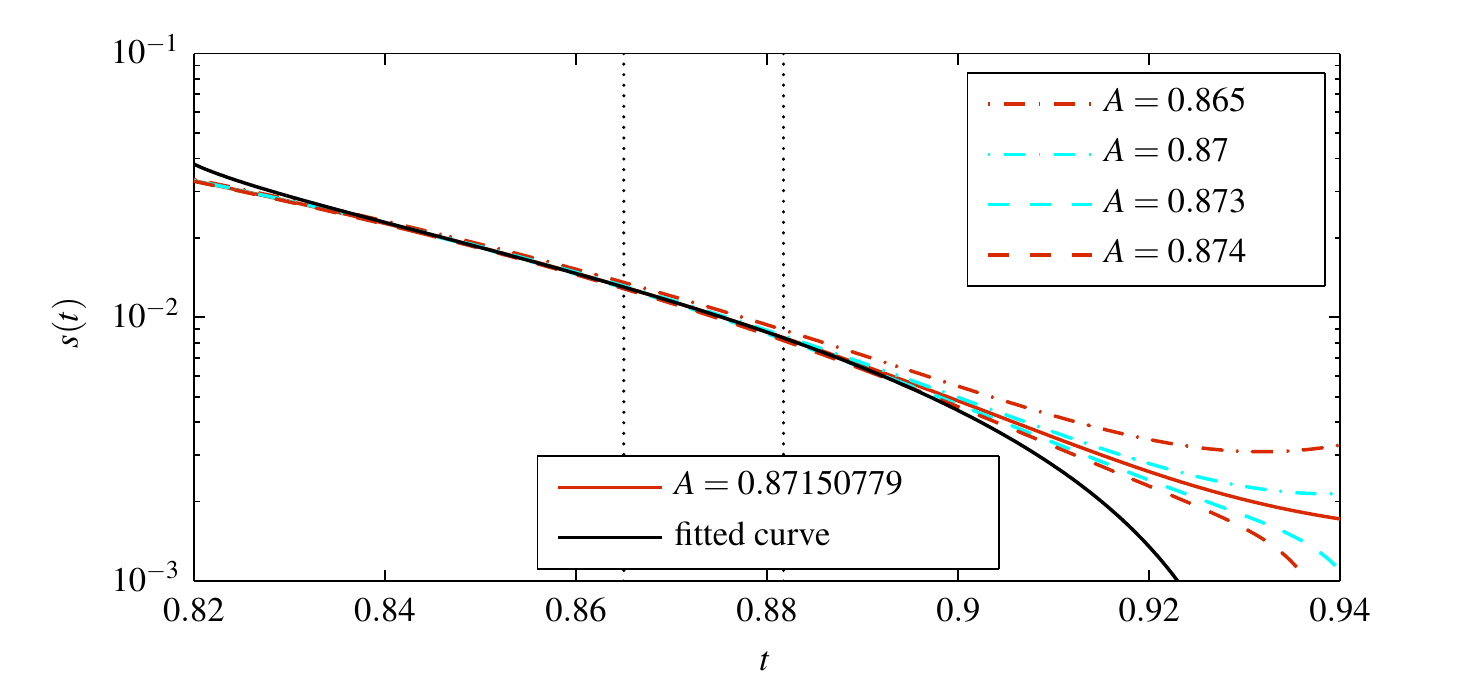}
    \caption{Fit of the scaling function $s(t)$ to the analytical expression, 
       given in \cite{OS2011}. The computed blow-up is $T = 
       0.94094524$. The value $A = 0.87150779$ is the last
       under-critical value shown, i.e., for which the solution does
       not change its behaviour at the origin. Computed with $N = 1281$.}
    \label{fig:ScalingFunctionFit_TC1_NonEquiv_QG_N1281}
  \end{figure}

  The next step is to determine the blow-up time $T$. This can be
  found by fitting the 
  last sub critical scaling function to the analytic expression (see~\cite{OS2011})
  \begin{align}\label{eq:scaling_func_analytic}
    s(t) = \frac{1.04}{\expo}\,(T-t)\,\exp\Bigl(-\sqrt{-\ln(T-t) + b}\Bigr)\,.
  \end{align}
  This formula was derived for the equivariant case but if the blow-up
  dynamics is a stable phenomenon then close to blow-up this formula
  should apply to the non-equivariant case as well. We fit the 
  curve for $A = 0.87150779$ to function \eqref{eq:scaling_func_analytic}.  This 
  results in $T = 0.93485135$ for the blow up time and $b = -2.1435346$ for the 
  parameter which depends on the initial data. The residual error for this fit was 
  $8.0327838\cdot 10^{-9}$. Figure \ref
  {fig:ScalingFunctionFit_TC1_NonEquiv_QG_N1281} shows the result of the fitting 
  procedure. The fit interval $t\in[0.865,0.8816]$ was chosen as a compromise 
  between being close enough to the blow-up time (lower time bound) and the time 
  domain, where the scaling function numerically converges (upper time bound).
  
\begin{figure}[hb]
    \centering
    \includegraphics{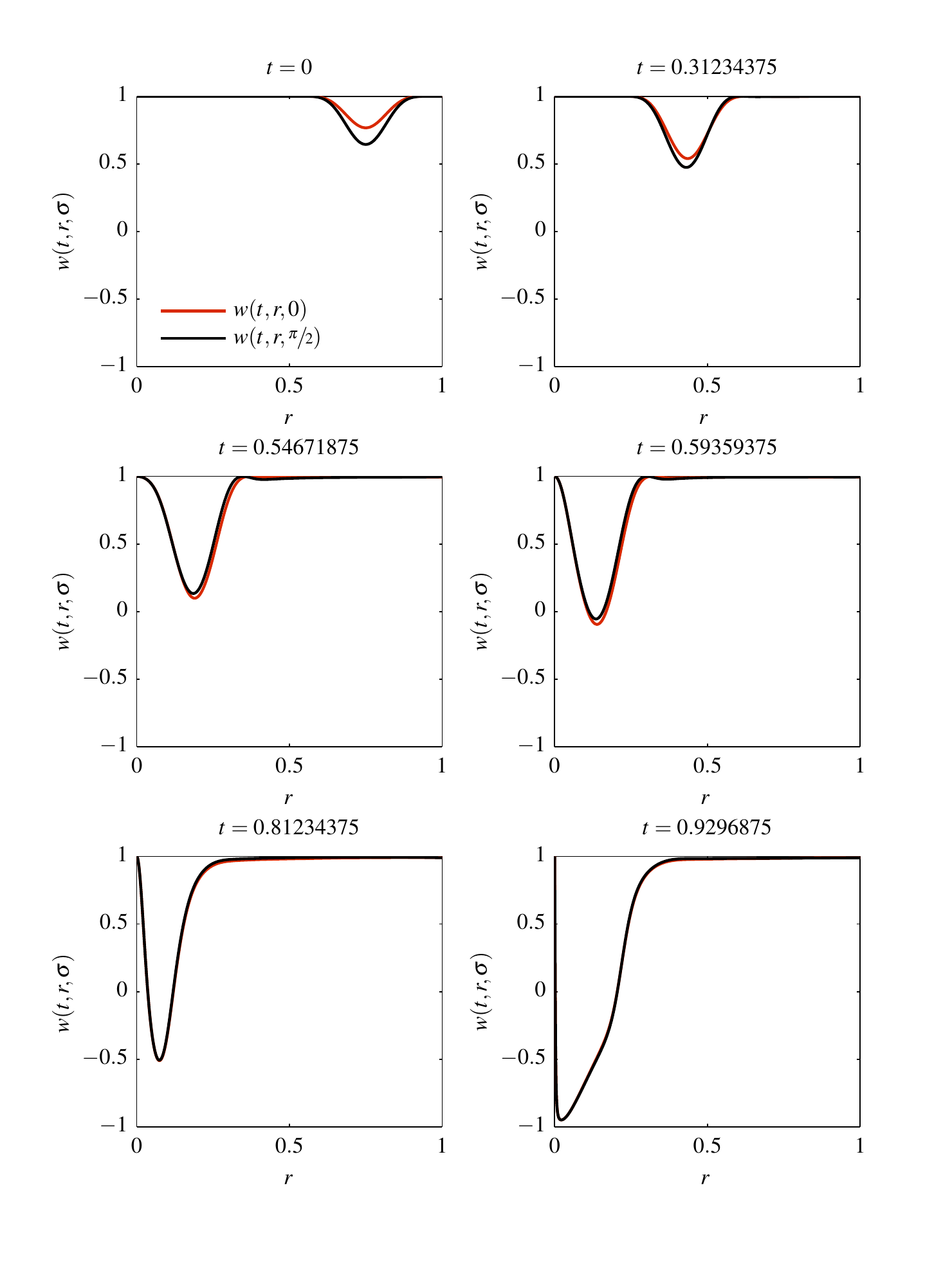}
    \caption{Time evolution of the wave packet for the amplitude $A = 
    0.87150779$. The slice along the $x$-axis and along the diagonal $y=x$ is 
    shown.}
    \label{fig:NonEquiv_EvolutionPlots}
  \end{figure}

\begin{figure}[hb]
    \centering
    \includegraphics{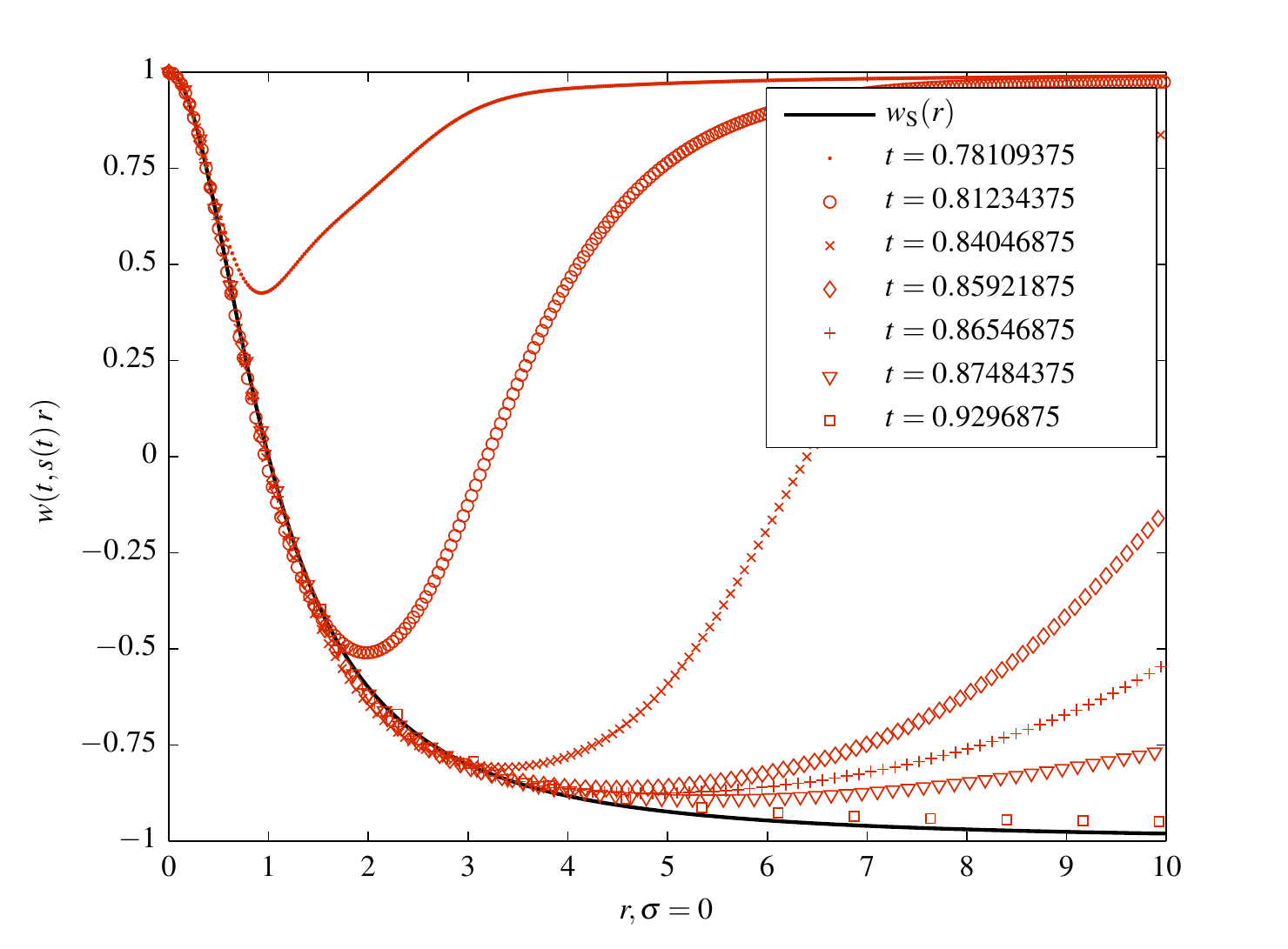}
    \caption{Ingoing wave packet for $A = 0.87150779$ and $B = 0.8$ rescaled 
       with the scaling function $s(t)$ for various values of $t$ along the
       $x$-axis. The
       rescaled functions $w(t,s(t) r)$ approximate the static
       solution $w_{\text{S}}(r)$. The function $w(t,r)$ reaches its
       minimum at $t = 0.9296875$.}
    \label{fig:w_slice_x_A08_N1281_TC1_NonEquiv}
  \end{figure}

  We present now, how the rescaled dynamic solution approximates the 
  equivariant harmonic map near the blow-up. Since equivariance is
  broken this process is not isotropic anymore but instead depends on
  the angle $\sigma$. In figure \ref{fig:NonEquiv_EvolutionPlots} we
  show the graphs of the component $w(t,x,y)$ taken along the $x$-axis
  and along the diagonal. As one can clearly see, as time progresses
  the profiles agree increasingly better. This being the case, we show
  in figure \ref{fig:w_slice_x_A08_N1281_TC1_NonEquiv} the successive stages
  of the rescaled dynamical solution taken only along the $x$-axis. 
  
  To 
  measure the deviation from the equivariant case, the difference
  between the two time steps when each of the wave packets reach their minimum 
  is used. Additionally, the difference in the minima itself is used.
  Table \ref{tbl:NonEquiv_WP_results} shows the numerical results of the two 
  rescaled wave packets. The time $t_{\text{min}}$ is defined as the time when 
  the wave packet along the $x$-axis respectively the line $y=x$, reaches its 
  minimum 
  $w_{\text{min}}(t_{\text{min}})$. Based on those results, the relative 
  deviation between the respective values are computed. Additionally, 
  the respective minima $w_{\text{min}}(0)$ at $t=0$ are shown.
  \begin{table}[h]
    \begin{center}
    \begin{tabular}{l c c c}
      &$x$-axis &diagonal &rel. deviation\\
      \hline
      $t_{\text{min}}$ &$0.9296875$ &$0.929375$ &$3.3624748\cdot 10^{-4}$\\
      $w_{\text{min}}(t_{\text{min}})$ &$-0.94862286$ &$-0.94867828$
                                       &$5.8418118\cdot 10^{-5}$\\
      $w_{\text{min}}(0)$ &$0.76663899$ &$0.64367501$ &$0.19103426$
    \end{tabular}
    \end{center}
    \caption{Comparison of the ingoing wave packet along the $x$-axis and 
       the diagonal}
    \label{tbl:NonEquiv_WP_results}
  \end{table}
  
  From the results in table \ref{tbl:NonEquiv_WP_results} and the
  graphs we conclude that the difference between the profiles  along the $x$-axis and $y=x$ vanishes during the
  evolution. The deviation along the two lines becomes significantly
  smaller, ending in nearly rotational symmetry. 
  
  
\section{Conclusions}
We presented here for the first time numerical indications for a
blow-up and singularity formation in the non-equivariant $2+1$
dimensional wave map system. The methods which were developed for the
analysis of the equivariant case of this system were extended and
generalised to the non-equivariant case.  It was possible to show that
using initial data close to equivariance can also to a blow-up
behaviour lead in the non-equivariant case.
  
  However, our simulations also showed the need for higher numerical
  resolutions in the 2-dimensional non-equivariant case
  to get a deeper and more detailed view into the blow-up dynamics and 
  singularity formation. This can be done with grid refinement techniques or by 
  changing the spatial discretisation of the equations completely. The use of 
  (pseudo) spectral methods could be an appropriate way.
%
%

%
\end{document}